\documentclass[5p,times,authoryear]{elsarticle}

\biboptions{square,comma}
\setcitestyle{numbers}

\usepackage[colorlinks=true,linkcolor=blue,citecolor=blue,urlcolor=blue]{hyperref}
\usepackage{graphicx}
\usepackage{dcolumn}
\usepackage{bm,ulem}
\usepackage{color}
\usepackage{hyperref}
\usepackage{soul}
\usepackage{amsmath}
\usepackage{amssymb}

\begin{document}

\tnotetext[t1]{LA-UR-25-31265}

\title{Constraining Hamiltonians from chiral effective field theory with neutron-star data}

\author[1,2]{Cassandra~L.~Armstrong}
\ead{armst520@msu.edu}
\author[3]{Brendan~T.~Reed}
\author[4]{Tate Plohr}
\author[5]{Henrik Rose}
\author[3]{Soumi De}
\author[3]{Rahul~Somasundaram}
\author[3]{Ingo Tews}

\address[1]{Facility for Rare Isotope Beams, Michigan State University, East Lansing, Michigan 48824, USA}
\address[2]{Intelligence and Space Research Division, Los Alamos National Laboratory, Los Alamos, NM 87545, USA}
\address[3]{Theoretical Division, Los Alamos National Laboratory, Los Alamos, NM 87545, USA}
\address[4]{Los Alamos High School, Los Alamos, NM 87544, USA}
\address[5]{Institut f{\"u}r Physik und Astronomie, Universit{\"a}t Potsdam, Haus 28, Karl-Liebknecht-Str. 24/25, 14476, Potsdam, Germany}

\date{\today}

\begin{abstract}
Multi-messenger observations of neutron stars (NSs) and their mergers have placed strong constraints on the dense-matter equation of state (EOS). 
The EOS, in turn, depends on microscopic nuclear interactions that are described by nuclear Hamiltonians.
These Hamiltonians are commonly derived within chiral effective field theory (EFT).
Ideally, multi-messenger observations of NSs could be used to directly inform our understanding of EFT interactions, but such a direct inference necessitates millions of model evaluations.
This is computationally prohibitive because each evaluation requires us to calculate the EOS from a Hamiltonian by solving the quantum many-body problem with methods such as auxiliary-field diffusion Monte Carlo (AFDMC), which provides very accurate and precise solutions but at a significant computational cost.
Additionally, we need to solve the stellar structure equations for each EOS which further slows down each model evaluation by a few seconds.
In this work, we combine emulators for AFDMC calculations of neutron matter, built using parametric matrix models, and for the stellar structure equations, built using multilayer perceptron neural networks, with the \texttt{PyCBC} data-analysis framework to enable a direct inference of coupling constants in an EFT Hamiltonian using multi-messenger observations of NSs.
We find that astrophysical data can provide informative constraints on two-nucleon couplings despite the high densities probed in NS interiors.
\end{abstract}

\maketitle

\section{Introduction}

Neutron-star (NS) cores explore densities far larger than those in the center of atomic nuclei, providing unique opportunities to test theories of strong interactions~\cite{Lattimer:2004pg,Watts:2016uzu,Chatziioannou:2024jsr}.
The properties of NSs, such as their radii, masses, and tidal deformabilities, depend on microscopic interactions between nucleons at densities up to a few times the nuclear saturation density $n_0\simeq0.16$~fm$^{-3}$ and, possibly, between exotic degrees of freedom at higher densities.
Multi-messenger observations of NS properties can, therefore, directly inform our understanding of these interactions.

Modern nuclear interactions are described by Hamiltonians derived within chiral effective field theory (EFT), which provides a systematic expansion of nuclear forces in powers of the average nucleon momenta $p$ over a breakdown scale $\Lambda_b$~\cite{Epelbaum:2008ga,Machleidt:2011zz}.
Chiral EFT Hamiltonians depend on a set of low-energy couplings (LECs) which determine the strength of individual operator structures that appear at different orders in this expansion.
In the two-nucleon (NN) sector, these LECs are typically adjusted to NN scattering data by solving the quantum-mechanical scattering problem~\cite{Entem:2003ft,Epelbaum:2004fk,Gezerlis:2013ipa,Piarulli:2014bda,Reinert:2017usi,Entem:2017gor,Somasundaram:2023sup}. 
In the three-nucleon (3N) sector, the LECs are usually fit to ground-state properties of light nuclei or the triton $\beta$ decay~\cite{Navratil:2007we,Gazit:2008ma,Hebeler:2010xb,Lynn:2015jua,Piarulli:2017dwd,LENPIC:2018ewt,Tews:2024owl,Curry:2025pna}, but different fit protocols have been proposed~\cite{Ekstrom:2015rta,Arthuis:2024mnl}.
The chiral EFT Hamiltonians are then employed in predictive calculations of larger quantum many-body systems, such as medium-mass to heavy atomic nuclei~\cite{Stroberg:2019bch,Miyagi:2021pdc,Hu:2021trw,Hebeler:2022aui,Miyagi:2023zvv,Door:2024qqz,Arthuis:2024mnl,Bonaiti:2025bsb,Li:2025exk,Kuske:2025tsm} or dense nuclear matter~\cite{Hebeler:2009iv,Tews:2012fj,Lynn:2015jua,Drischler:2017wtt,Drischler:2020yad,Lovato:2022apd,Keller:2022crb,Tews:2024owl,Marino:2024tfp}.
\begin{figure*}[t]
    \centering
    \includegraphics[height=0.6\columnwidth]{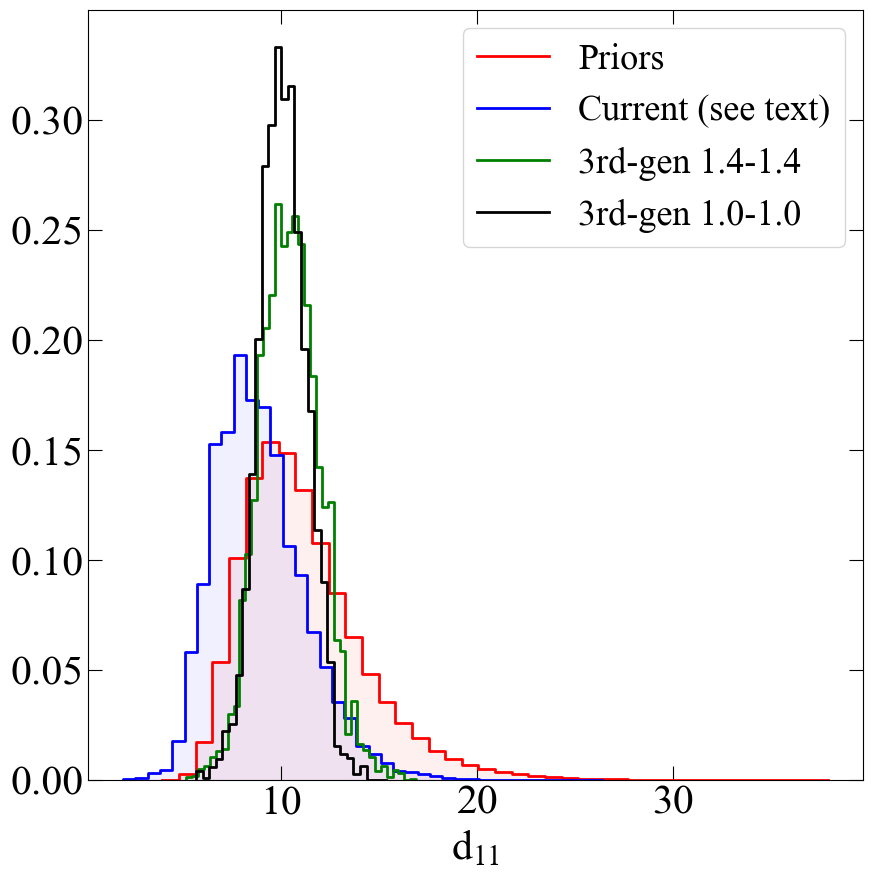}
    \includegraphics[height=0.6\columnwidth]{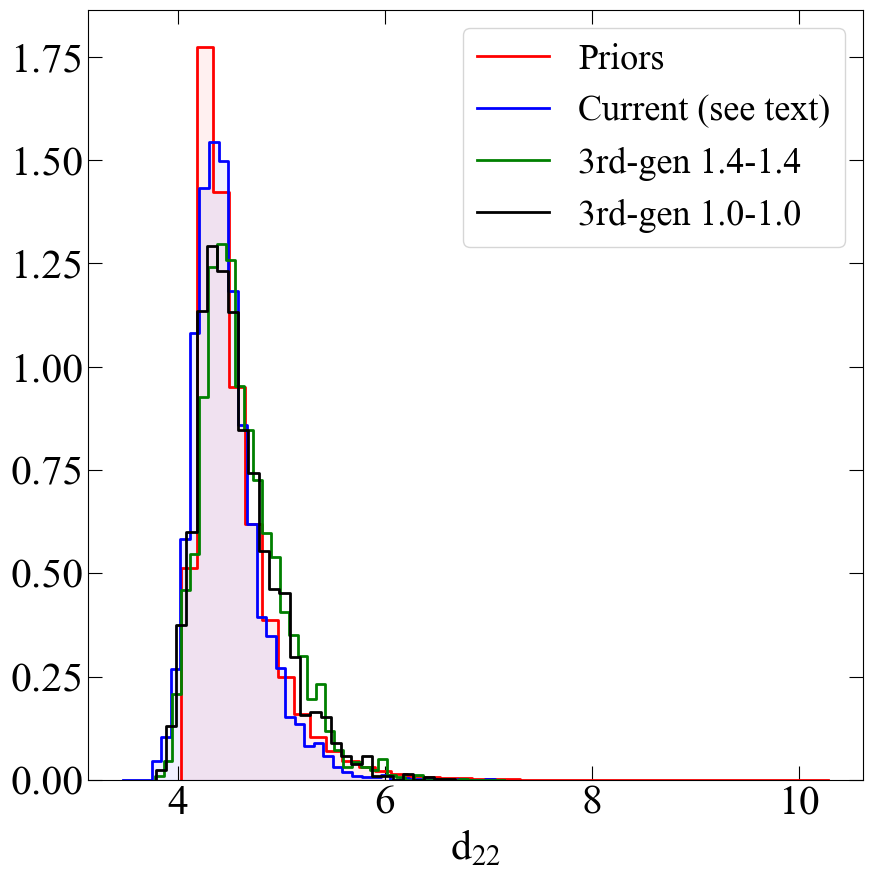}
    \includegraphics[height=0.6\columnwidth]{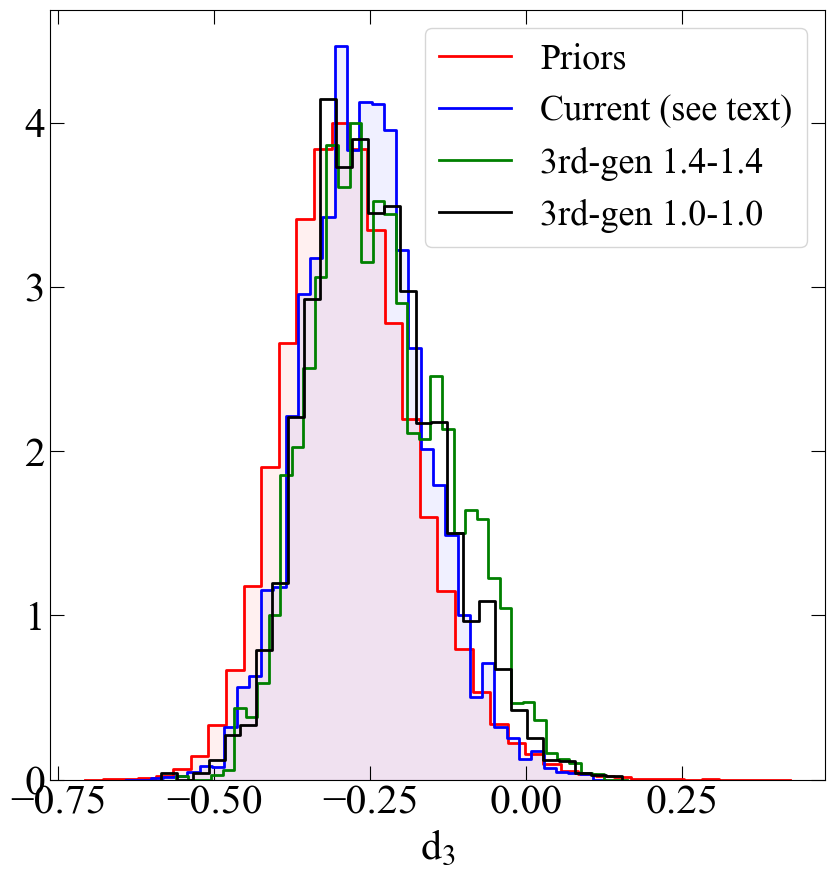}
    \includegraphics[height=0.6\columnwidth]{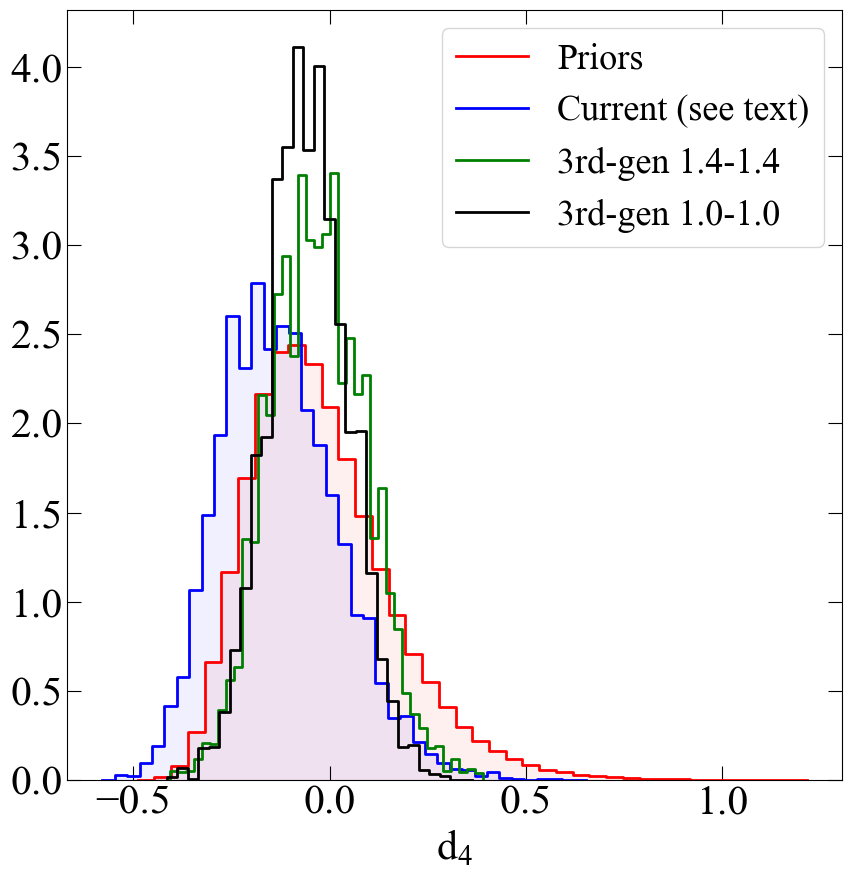}
    \includegraphics[height=0.6\columnwidth]{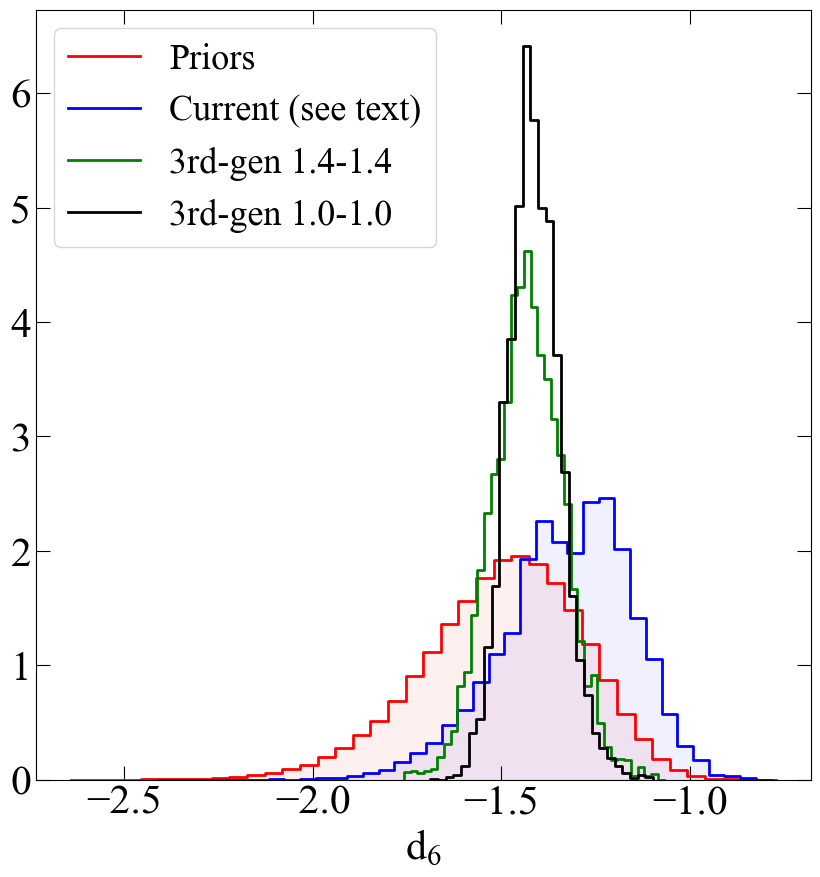}
    \includegraphics[height=0.6\columnwidth]{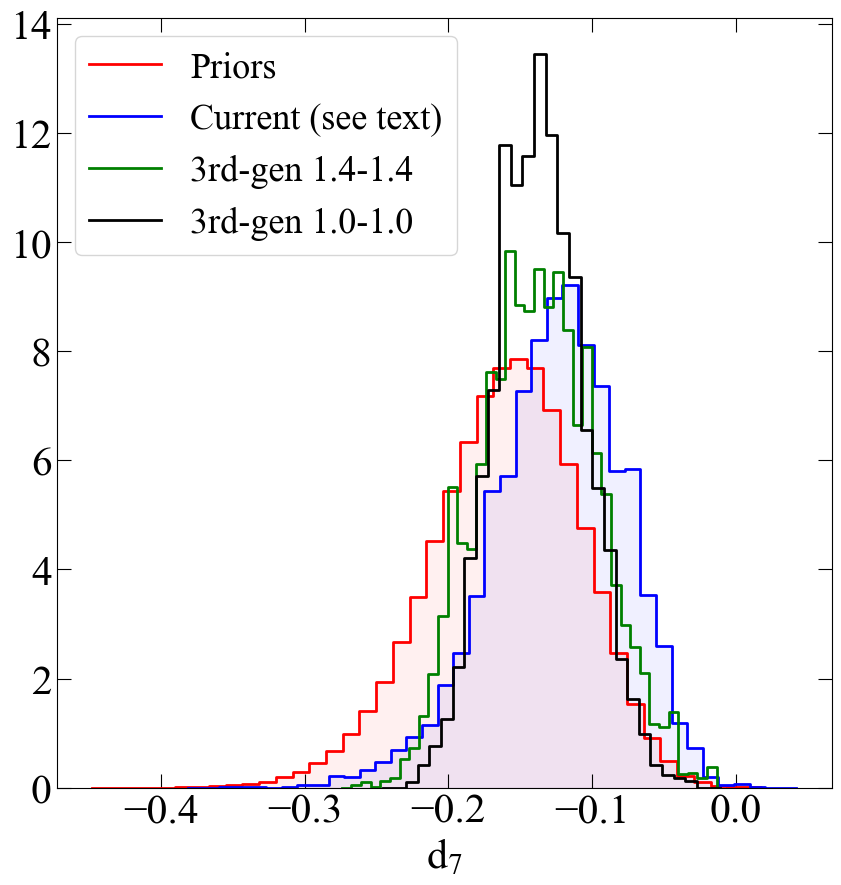}
    \caption{Priors (red) and posteriors of the analysis of astrophysical data for all six spectral LECs in neutron matter. 
    The priors are results of fits to NN scattering data for a single cutoff of $R_0=0.6$~fm. 
    The posterior for current astrophysical data includes gravitational-wave, maximum-mass, and NICER data (blue).
    We also show posteriors from analyses of simulated signals as expected to be observed by third-generation gravitational-wave detectors for symmetric binaries with component masses of $1.0M_{\odot}$ (black) and $1.4M_{\odot}$ (green).
    }
    \label{fig:lecs}
\end{figure*}
Terrestrial nuclear systems, however, only explore small neutron-to-proton asymmetries and densities of the order of $n_0$.
Matter in NSs, in contrast, is very isospin-asymmetric and probes much larger densities. 
Adjusting nuclear forces and their LECs to properties of NS matter would provide valuable complementary constraints~\cite{Maselli:2020uol,Sabatucci:2022qyi,Rose:2023uui,Somasundaram:2024ykk}. 
In this work, we explore how well data from NS mass measurements~\cite{Antoniadis:2013pzd,NANOGrav:2017wvv,Fonseca:2021wxt}, from a gravitational-wave (GW) observation of two merging NSs, GW170817~\cite{LIGOScientific:2017vwq}, and from the Neutron Star Interior Composition Explorer (NICER) telescope~\cite{Vinciguerra:2023qxq,Salmi:2024aum,Choudhury:2024xbk,Mauviard:2025dmd} enable us to constrain the relevant LECs in a nuclear Hamiltonian from chiral EFT at next-to-next-to-leading order (N$^2$LO) in neutron matter.

\section{Approach}

To extract LECs from NS observations, our goal is to employ Bayesian inference to analyze astrophysical data.
Bayesian inference necessitates millions of model evaluations, and hence, it requires us to be able to rapidly connect LECs with NS observables.
To date, this connection is too computationally expensive.
First, it requires solving the quantum many-body problem to obtain the dense-matter EOS from a given set of LECs, a process which can cost 100,000's of CPU hours per EOS.
Second, NS observables are calculated from an EOS by solving the Tolman-Oppenheimer-Volkoff (TOV) equations, which takes a few seconds per EOS.
These two steps are prohibitively slow for Bayesian inference; typically, likelihood evaluations need to be performed in less than $\sim1s$~\cite{Reed:2024urq}.
Therefore, reducing this computational burden is essential.

Recent works have taken first steps in this direction.
Refs.~\cite{Maselli:2020uol,Sabatucci:2022qyi} obtained constraints on a simplified phenomenological Hamiltonian with a single free parameter representing the strength of a short-range 3N force from NS observations.
This one-parameter model described the EOS over the whole density range in a NS.
However, a connection to 3N forces from EFT and the possible existence of non-nucleonic degrees of freedom at high densities were not addressed in that work. 
This was overcome in Ref.~\cite{Somasundaram:2024ykk} where the two pion-nucleon LECs that determine the strength of chiral 3N forces in neutron matter at N$^2$LO were studied using a more flexible EOS model. 
In these works, the NN sector of the nuclear interaction was frozen, broad uniform prior distributions on the 3N LECs were employed, and canonical constraints from nuclear experiments, such as NN scattering data analyzed consistently with EFT truncation uncertainties, were not taken into account.

Here, we overcome these limitations and study the impact of NS observations on all unknown LECs in an N$^2$LO Hamiltonian in neutron matter, rather than limiting ourselves to the 3N sector. 
We achieve this by combining our previously developed emulators~\cite{Armstrong:2025tza} for auxiliary-field diffusion Monte Carlo (AFDMC)~\cite{Carlson:2014vla,Lynn:2019rdt} calculations of neutron matter and for the TOV equations~\cite{Reed:2024urq} to establish a direct connection of the LECs at N$^2$LO with NS properties. 
As the long-range pion-nucleon LECs are very tightly constrained from experimental data~\cite{Hoferichter:2015hva,Siemens:2016jwj}, we keep them fixed in this work. 
This approximation is further justified as these LECs can only be meaningfully measured with about a dozen next-generation detections~\cite{Somasundaram:2024ykk}, a scenario we do not study here.
We also employ a local N$^2$LO Hamiltonian at a single cutoff of $R_0=0.6$~fm~\cite{Somasundaram:2023sup}, because at this cutoff we find regulator artifacts to be negligible~\cite{Tews:2024owl}. 
We do not explore smaller cutoff values because we would need to explicitly account for spurious shorter-range 3N contributions~\cite{Lynn:2015jua}, increasing the computational cost of training emulators and making our results less robust.
With these choices, the 3N sector in neutron matter is parameter-free at this order in chiral EFT because the 3N contact interaction vanishes due to the Pauli principle and the one-pion--exchange-contact interaction vanishes due to its spin-isospin structure~\cite{Hebeler:2009iv}.
Consequently, the contribution of 3N forces to neutron matter in chiral EFT is completely predicted from pion-exchange terms alone, which remains true even for subleading 3N forces~\cite{Tews:2012fj}.
Hence, we effectively study the contact sector of the NN force. 
This allows us to analyze the interplay between NN scattering data and NS observations in constraining the Hamiltonian. 

We employ both of our emulators to sample over the LECs when analyzing NS data and performing GW inference with \texttt{PyCBC}~\cite{Reed:2025sqh}.
Prior information from NN scattering data is incorporated in a hierarchical manner within \texttt{PyCBC} by implementing LEC posteriors from a scattering analysis~\cite{Somasundaram:2023sup} as GW priors.
We present our main results in Fig.~\ref{fig:lecs}, and find that present astrophysical data provide slight constraints on the relevant spectral LECs at N$^2$LO if nuclear Hamiltonians are used up to $2n_0$. 
We also study the potential of next-generation GW detectors and find that these observatories can provide strong constraints on LECs, highlighting their high discovery potential for nuclear physics.

\begin{figure}[t]
    \centering
    
\begin{tabular}{ccc}
        \includegraphics[width=0.8\columnwidth]{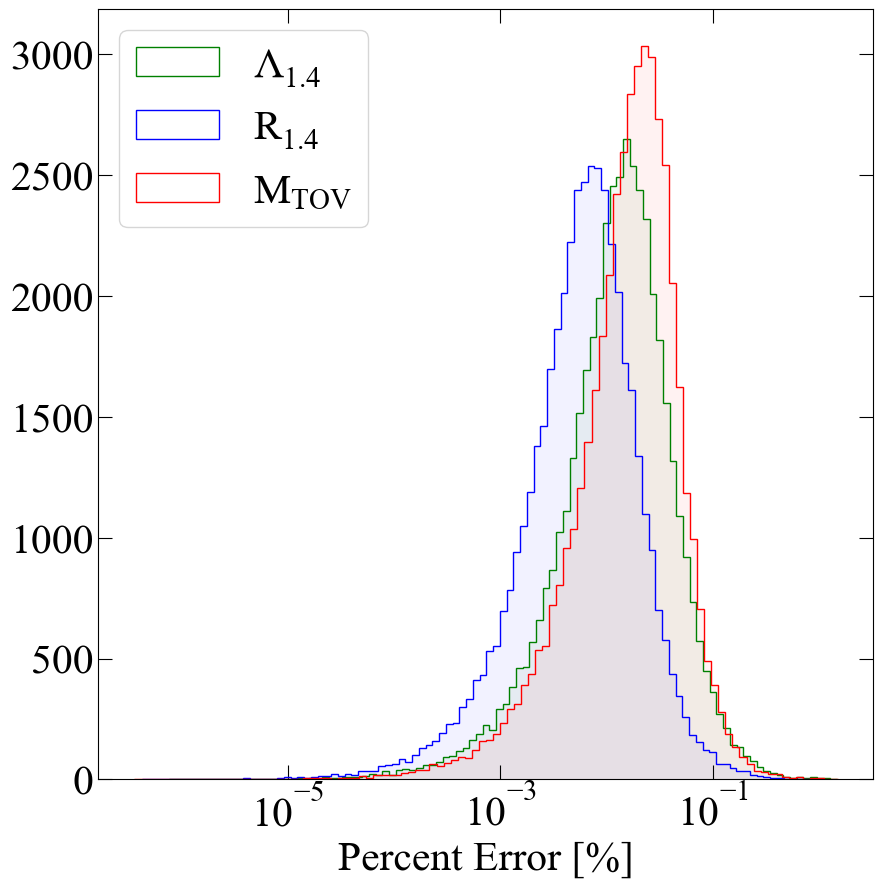} 
    \end{tabular}
    
    \caption{Percent error of the MLP neural network emulators for  M$_{\rm TOV}$ (red), the radius (blue),  and the tidal deformability (green) of a $1.4$M$_{\odot}$ neutron star.
    In all panels, we show histograms for 70,000 validation samples.
    }
    \label{fig:emulator_error}
\end{figure}

\section{Emulators for Dense Matter and Neutron Star Structure}

For a given Hamiltonian with a specified set of LECs, we employ AFDMC to calculate the EOS of pure neutron matter.
AFDMC provides very accurate and precise solutions to the quantum many-body problem but at a significant computational cost.
We have employed the parametric matrix model (PMM)~\cite{Cook:2024toj} to build emulators for AFDMC calculations of pure neutron matter for local chiral EFT Hamiltonians at N$^2$LO~\cite{Armstrong:2025tza}.
The PMM is based off of the reduced-basis method (RBM)~\cite{Frame:2018,Konig:2020,Bonilla:2022,Drischler:2022ipa,Melendez:2022kid,Duguet:2023wuh,Giuliani:2023} which identifies a relevant low-dimensional subspace that contains the solution space as the LECs in the Hamiltonian are varied.
While the RBM explicitly calculates this subspace from high-fidelity solutions to the problem at hand, the PMM identifies this subspace implicitly through fitting matrix elements in a representative low-dimensional matrix equation directly to the AFDMC data.
This equation is set up as an eigenvalue problem for a matrix $\hat{H}$ with the same structure as the chiral EFT Hamiltonian:
\begin{equation}
\hat{H} = H_0 + d_1 \cdot H_1+d_2 \cdot H_2+\cdots\,,
\label{eq:PMM}
\end{equation}
where the $d_i$ are the LECs of the Hamiltonian, and e.g., in the 2-dimensional case, 
\begin{align}
    \hat{H}&=\begin{bmatrix}
                    \alpha & \beta \\
                    \beta & \gamma
                \end{bmatrix}, 
                H_0=\begin{bmatrix}
                    a & 0 \\
                    0 & b
                \end{bmatrix}, \nonumber \\
    H_1&=\begin{bmatrix}
                    c & d \\
                    d & e
                \end{bmatrix},
                H_2=\begin{bmatrix}
                    f & g \\
                    g & h
                \end{bmatrix}\,,
                \label{eq:PMM_matrix}
\end{align}
with real matrix elements $a-h$.
These matrix elements are constrained by fitting the lowest eigenvalue of $\hat{H}$ to AFDMC energies in neutron matter for $\sim 30$ training LEC sets. 
At N$^2$LO, there are nine operator LECs describing NN forces, but only six combinations of these operator LECs, i.e. spectral LECs, contribute to pure neutron matter.
The emulator is set up with the six relevant spectral LECs $d_i$.
As explained before, the 3N forces are fixed because, in neutron matter at the chosen cutoff value, they only depend on pion-nucleon LECs at N$^2$LO.
The emulators developed in Ref.~\cite{Armstrong:2025tza} allow us to speed up AFDMC calculations by a factor of $10^8$ while maintaining errors around 1\%, which is comparable to the statistical uncertainty of AFDMC~\cite{Carlson:2014vla}.
This dramatic reduction of computational cost enables us to perform Bayesian inference.
More details on our PMM emulator can be found in Ref.~\cite{Armstrong:2025tza}.

For a given set of LECs, the PMM allows us to calculate the neutron-matter energy per particle at 0.12, 0.16, 0.24, 0.32, 0.48~fm$^{-3}$.
We then use the nuclear meta-model~\cite{Margueron:2017eqc,Margueron:2017lup} to extend the PMM output to neutron-star matter.
The meta-model describes the energy per particle of nuclear matter at various densities and asymmetries in terms of the nuclear empirical parameters (NEPs), defined by Taylor expansions of the energy per particle in symmetric nuclear matter and of the symmetry energy around nuclear saturation density.
We first generate a set of symmetric-matter parameters.
We fix $E_{\rm sat} = -16$~MeV and $n_{\rm sat} = 0.16$~fm$^{-3}$, and sample $K_{\rm sat}$, $Q_{\rm sat}$, and $Z_{\rm sat}$ from the following Gaussian distributions: $K_{\rm sat}=227\pm18$~MeV, $Q_{\rm sat}=-172\pm243$~MeV, and $Z_{\rm sat}=1287\pm1499$~MeV~\cite{Somasundaram:2020chb}.
We then sample a set of six spectral LECs $d_{11}$, $d_{22}$, $d_{3}$, $d_{4}$, $d_{6}$, and $d_{7}$ from the posteriors of Ref.~\cite{Somasundaram:2023sup}, and use the PMM to generate a neutron-matter EOS. 
With both curves specified, we fit the meta-model, and generate an EOS in beta equilibrium. 
For all models, we attach the crust of Douchin and Haensel~\cite{Douchin:2001sv} below 0.08~fm$^{-3}$ following the approach outlined in Ref.~\cite{Koehn:2024set}.
Therefore, each meta-model EOS depends on a set of 9 parameters.

Because we cannot expect the nucleonic description of dense matter to be valid at the highest densities reached in the cores of NSs, we transition from the meta-model to a speed-of-sound extension~\cite{Tews:2018iwm,Greif:2018njt,Somasundaram:2021clp} at a transition density $n_{\rm tr}=2n_0$.
We sample values of the squared speed of sound, $c_i^2$, uniformly between $10^{-5}$ and 1 on a density grid ranging from $3n_0$ to $6n_0$ in steps of $n_0$.
We then connect these points by linear speed-of-sound segments and choose the speed of sound to be constant after the last density point.
Finally, we connect the speed-of-sound model to the meta model and construct the EOS at all densities up to $12n_0$.
Each final EOS is, therefore, described by a total of 13 parameters.
We have tested that increasing the number of speed-of-sound parameters does not significantly affect our results, which is in agreement with previous findings~\cite{Somasundaram:2024ykk}.
This is because the inference presented in this work is mostly sensitive to properties of light to medium-mass NSs which do not probe the EOS at the hightest densities.

To build emulators for the TOV equations, we follow Ref.~\cite{Reed:2024urq} and employ multilayer-perceptron (MLP) neural networks.
We generate mass--radius and mass--tidal-deformability curves for 270,000 EOS, and split these into a training set (200,000 EOS) and a validation set (70,000 EOS).
We remove all samples with a maximum NS mass M$_{\rm TOV}~<~2$M$_{\odot}$ because the observations of heavy pulsars with masses greater than $2$M$_{\odot}$ rule these EOS out.
We then build three separate MLP networks to respectively emulate M$_{\rm TOV}$, the radii $r(m)$, and the tidal deformabilities $\Lambda(m)$.
Each MLP emulator takes the 13 parameters describing each EOS as input.
To emulate M$_{\rm TOV}$, we train an MLP regressor to the M$_{\rm TOV}$ of our training EOS.
For the validation set, our emulator predicts M$_{\rm TOV}$ with errors of $\sim 0.01$\%, see Fig.~\ref{fig:emulator_error}.
To build emulators for the radii and tidal deformabilities, we define a mass grid of 30 uniformly spaced points between 1M$_{\rm \odot}$ and M$_{\rm TOV}$ for each of the training EOS.
Two MLP regressors are then trained on the radii and the logarithm of the tidal deformabilities at the grid points, respectively.
For any new parameter set, we first predict M$_{\rm TOV}$ and then use the other emulators to infer radii and tidal deformabilities at the mass-grid points.
We find that these emulators perform extremely well, with speed ups of two orders of magnitude~\cite{Reed:2024urq} and errors around 0.01\% over the validation set, see Fig.~\ref{fig:emulator_error} for the error distributions for both quantities at 1.4M$_{\rm \odot}$. 
We show example outputs in the Supplemental Material.

\section{Multimessenger Data Analysis}

Recent years have seen a wealth of new NS data. 
Radio observations of heavy pulsars have firmly established the existence of two-solar-mass NSs~\cite{Antoniadis:2013pzd,NANOGrav:2017wvv,Fonseca:2021wxt}.
The first gravitational-wave (GW) observation of a binary NS merger, GW170817~\cite{LIGOScientific:2017vwq}, constrained the tidal deformability for a $1.4$M$_\odot$ NS to be $190^{+390}_{-120}$ at the 90\% credible level~\cite{LIGOScientific:2018cki}, implying the EOS to be rather soft up to $(3-4)n_0$.
Finally, NICER has measured the masses and radii of four pulsars~\cite{Vinciguerra:2023qxq,Salmi:2024aum,Choudhury:2024xbk,Mauviard:2025dmd}.

To analyze GW170817 data, we follow Ref.~\cite{Reed:2025sqh} to implement the MLP networks in the \texttt{PyCBC} GW Bayesian analysis software. 
We compute the posterior $p(\vec\theta|d(t),H)$ by sampling over the parameters $\vec\theta$ that enter the waveform model $H$, for which we choose the SEOBNRv4T\_surrogate template~\cite{pratten_2025_14999310}. 
The parameters $\vec\theta$ are the 13 parameters that describe the EOS as well as the parameters describing the binary system.
For the EOS parameters, we use our emulators to predict $r(m)$ and $\Lambda(m)$ between $1.0$M$_{\odot}$ and M$_{\rm TOV}$.
As in Ref.~\cite{Reed:2025sqh}, we fix the sky and distance parameters to the ones for GW170817~\cite{DES:2017kbs,Cantiello:2018ffy} and sample the source frame chirp mass $\mathcal{M}_c^{\rm src}$, mass ratio $q$, offset of trigger time $\Delta t_c$, inclination angle $\iota$, and the two NS spins $\chi_{1z,2z}$, and marginalize the polarization angle and phase using a uniform prior from $[0,2\pi)$ for both.
We compare the predicted waveform to the strain data $d(t)$ measured by the LIGO Hanford, LIGO Livingston, and Virgo detectors~\cite{Vallisneri:2014vxa,LIGOScientific:2019lzm}, by employing \texttt{PyCBC} Inference~\cite{Biwer:2018osg} and sample using the \textsc{emcee} Markov Chain Monte Carlo (MCMC) sampler. 
The inference calls the emulator via a plug-in waveform model developed in Ref.~\cite{Reed:2025sqh}.

\begin{figure}[t]
    \centering
    \includegraphics[height=0.87\columnwidth]{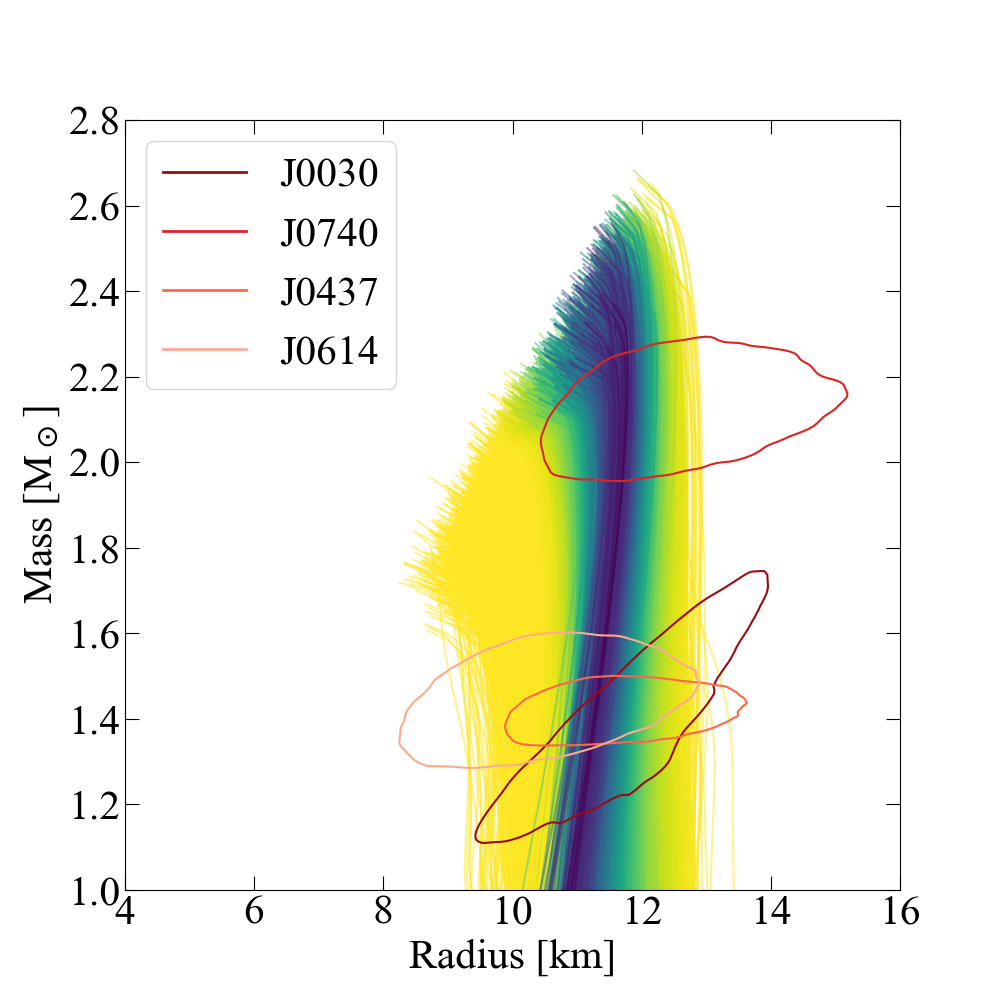}
    \includegraphics[height=0.88\columnwidth]{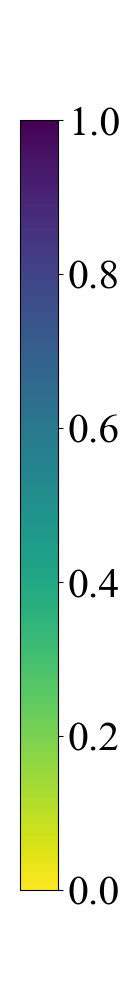}
    \caption{Mass-radius curves drawn from the GW posterior where the maximum-mass and NICER likelihoods are applied. 
    The EOS are colored according to their posterior probability normalized to the highest value.
    We also show the four NICER contours at 95\% credible level. 
    }
    \label{fig:mr_inference_likelihood}
\end{figure}

In Fig.~\ref{fig:mr_inference_likelihood}, we
show 9800 mass-radius relations drawn from the GW posterior of our {\texttt{PyCBC}} analysis.
The main effect of the GW data is to rule out stiffer EOS, consistent with previous analyses~\cite{LIGOScientific:2017vwq,Annala:2017llu,Capano:2019eae,Dietrich:2020efo,Essick:2021kjb,Miller:2021qha,Raaijmakers:2021uju,Rutherford:2024srk,Somasundaram:2024ykk}.
We further apply the maximum-mass constraint by comparing the predicted $M_{\rm max}$ for each EOS to the observed neutron star masses~\cite{Koehn:2024set}. 
The likelihood is dominated by PSR J0740+6620~\cite{Fonseca:2021wxt} with a mass of $2.08\pm0.07\, M_{\odot}$.
Finally, we apply mass-radius constraints from the four NICER pulsars PSR J0030+0451~\cite{Vinciguerra:2023qxq}, PSR J0740+6620~\cite{Salmi:2024aum}, PSR J0437–4715~\cite{Choudhury:2024xbk}, and PSR J0614-3329~\cite{Mauviard:2025dmd}.
For the first three pulsars, independent results have been provided by the Maryland group of the NICER collaboration~\cite{Miller:2019cac,Miller:2021qha,Miller:2025qfq}.
We compare our findings using results from both groups in the Supplemental Material.
Given all sources of data, we can put tight constraints on the NS mass-radius relation. 
We find the radius of a $1.4 M_{\odot}$ NS to be $11.6^{+0.7}_{-0.5}$~km and the pressure at twice saturation density to be $13.4^{+5.8}_{-6.0}$~MeV/fm$^3$, both at the 90\% credible level.

\section{Constraining Nuclear Interactions}

Because we sample over the EOS parameters during the data analysis, we can directly infer the LECs from multi-messenger observations of NSs and their mergers.
We present our results for the LECs given present astrophysical observations in Fig.~\ref{fig:lecs} and find that current NS data provides modest information on some spectral LECs. 
While the distributions of the LECs that enter PNM through their contribution to singlet $S$-wave interactions ($d_{22}$ and $d_3$)~\cite{Gezerlis:2014zia} remain nearly unchanged, the distributions of the LECs entering the triplet $P$-wave channels ($d_{11}, d_4, d_6$, and $d_7$) undergo moderate change after the inference is performed.
The astrophysical data reduces $P$-wave repulsion, which is in line with NS data preferring softer EOS. 
This is driven by both GW and NICER data which pull the LEC posteriors in the same direction, see the Supplemental Material, where we also provide a correlation plot for the spectral LECs and the radius of a $1.4M_{\odot}$ NS.
As a consequence, even present NS observations provide information on NN forces in the nuclear Hamiltonian, beyond what is provided by scattering data. 

\begin{figure}[t]
    \centering
\begin{tabular}{c}
        \includegraphics[width=0.45\textwidth]{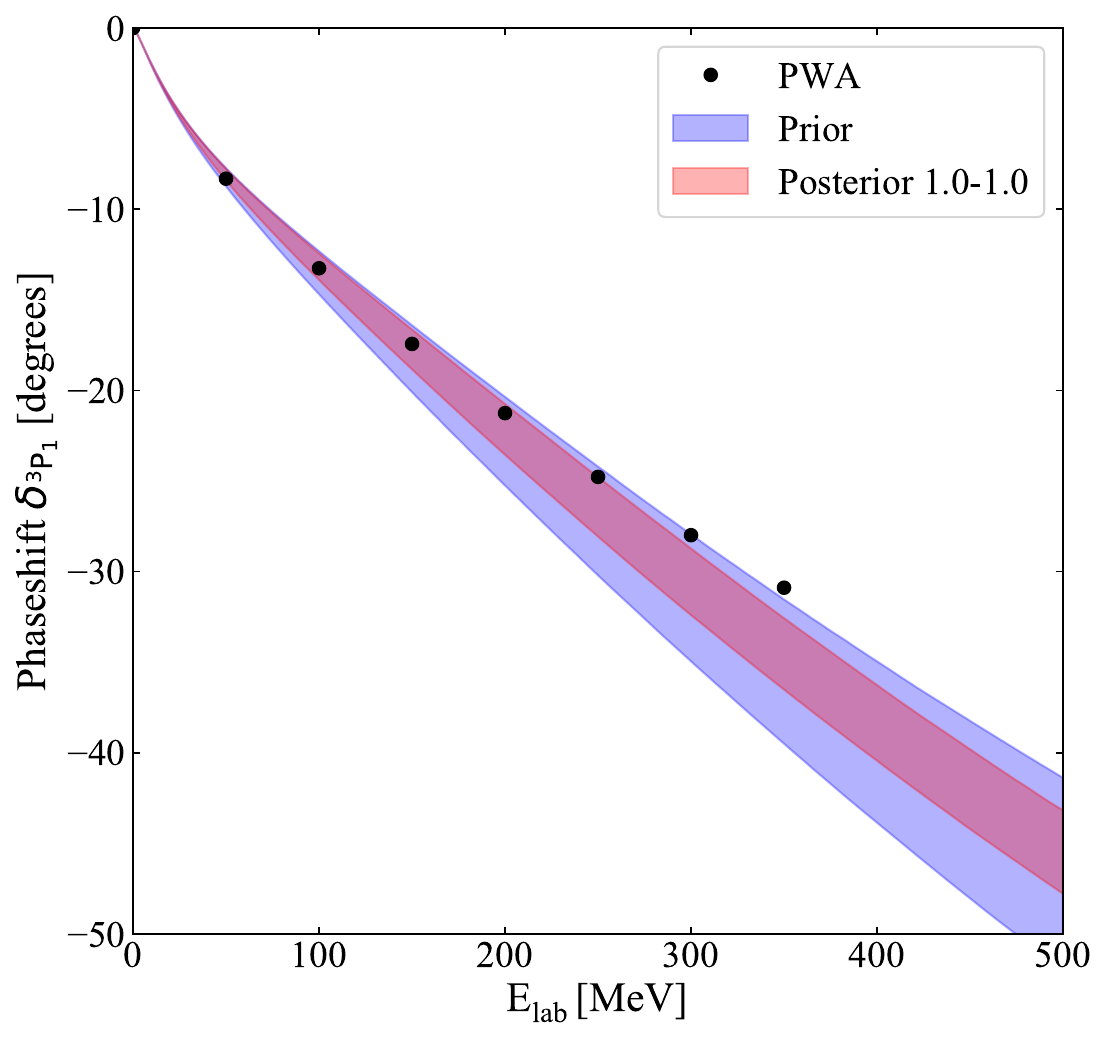} 
    \end{tabular}
    \caption{Phase shifts for the $^3P_1$ partial wave at the 90\% CI for the prior (blue) and the posterior obtained from the analysis of the $1.0-1.0$~$M_\odot$ injection (red) at a cutoff of $R_0=0.6$~fm.
    We also show the results of the Nijmegen partial-wave analysis~\cite{Stoks:1993tb} for comparison (black).}
    \label{fig:phase_shifts}
\end{figure}

Future observations, for example by next-generation GW detectors such as Cosmic Explorer~\cite{Reitze:2019iox,Evans:2021gyd} or the Einstein Telescope~\cite{Punturo:2010zz,ET:2019dnz}, will provide additional precision data that allows us to further constrain the LECs. 
To demonstrate this, we infer projected constraints on the LECs by injecting simulated events into a detector network composed of two Cosmic Explorer interferometers and one Einstein Telescope. 
Since the true distribution of these parameters for the potentially observable population of BNS systems is unknown, the details of the projected constraints will depend on the specific values chosen for the injection parameters. 
We explore this ambiguity by injecting two separate events which differ only in the chirp mass which primarily determines the density at which the EOS is probed by a GW signal.   
For both events, all other injection parameters, including the mass ratio, are chosen such that they are consistent with the inferred posteriors for GW170817.
For the first event, our choice for the chirp mass results in the two component masses being $1.4$~M$_\odot$, leading to an SNR of $\sim 2300$. 
For the second event, we decrease the chirp mass such that the two component masses are $1.0$~M$_\odot$, in order to probe the sensitivity of the LECs to the NS masses.
This event was found to have an SNR of $\sim 1730$. 
The injected LEC values were chosen to be close to the 50th percentiles of the corresponding prior distributions. 
The obtained posteriors for the LECs are also shown in Fig.~\ref{fig:lecs}. 
We find that next-generation GW detectors can provide significant constraints on the LECs for these two events. 
Interestingly, the LECs are better constrained for the lower SNR (low-mass) event as it probes lower densities that are closer to the regime described by chiral EFT.  
We expect that freeing the pion-nucleon LECs $c_i$ would have little impact on our findings~\cite{Somasundaram:2024ykk}.
Again, we note that varying the chirp mass in a broader range or choosing different values for the other injection parameters will change the quantitative details of the inferred posteriors, but we believe that such differences are captured at a qualitative level in our two-event injection study.  

Finally, for these next-generation posteriors, we re-examine the phase shifts in the triplet $P$-wave channels. 
We find no changes to the phase shift in the $^3$P$_0$ partial wave and marginal changes in the $^3$P$_2$ channel with respect to the priors.
However, we observe a clear change in the $^3$P$_1$ channel mainly at high energies $\gtrsim 200$~MeV, see Fig.~\ref{fig:phase_shifts}.
This channel is repulsive and leads to the largest of the $P$-wave phase shifts whereas $^3$P$_0$ and $^3$P$_2$ are both attractive and smaller in magnitude.
As we found before, the data seems to overall reduce $P$-wave repulsion.
This analysis serves as an important sensitivity study which demonstrates that, among the triplet $P$ waves, the $^3$P$_1$ channel is more strongly correlated with the properties of NSs. 
In addition, the fact that the spectral LECs change considerably after next-generation GW data is included while the phase shift bands change only moderately shows that astrophysical data provides complementary information when compared to NN scattering. 

\section{Summary and Outlook}

In this work, we have used emulators based on the PMM and MLP machine-learning frameworks to constrain the relevant two-nucleon LECs in an N$^2$LO chiral EFT Hamiltonian from NS data.
We found that current observations only provide mild constraints in addition to two-nucleon scattering measured in the lab, but that next-generation GW observatories provide strong and complimentary information on microscopic LECs.
Our pipeline paves the way of measuring LECs that are difficult to extract from laboratory data.
It can also be used to shed light on the limitations of interactions derived within chiral EFT~\cite{Somasundaram:2024ykk}.

\section*{Acknowledgements}
This work was supported by the U.S. Department of Energy through Los Alamos National Laboratory (LANL). 
LANL is operated by Triad National Security, LLC, for the National Nuclear Security Administration of U.S. Department of Energy (Contract No.~89233218CNA000001).
C.L.A. was supported by the Michigan State University Distinguished Fellowship (UDF) from the Michigan State University Graduate School. 
C.L.A., B.T.R., S.D., R.S., and I.T. were supported by the Laboratory Directed Research and Development (LDRD) program of LANL under project number 20230315ER.
C.L.A., R.S., and I.T. were also supported by the LDRD program of LANL under project number 20260260ER.
C.L.A. and B.T.R. were also supported by LANL through its Center for Space and Earth Science, which is funded by LANL’s LDRD program under project number 20240477CR.
B.T.R. and I.T. were also supported by the U.S. Department of Energy, Office of Science, Office of Nuclear Physics program under Award Number DE-SCL0000015. 
S.D. was also supported by the Laboratory Directed Research and Development program of Los Alamos National Laboratory project 20250750ECR.
Computational resources have been provided by the Los Alamos National Laboratory Institutional Computing Program, which is supported by the U.S. Department of Energy National Nuclear Security Administration under Contract No. 89233218CNA000001, and by the National Energy Research Scientific Computing Center (NERSC), which is supported by the U.S. Department of Energy, Office of Science, under Contract No. DE-AC02-05CH11231.

\bibliography{refs}



\clearpage
\onecolumn
\setcounter{figure}{0}
\setcounter{section}{0}
\setcounter{page}{1}

\section*{\Large Supplemental Material}

\section{Emulators for TOV}

\begin{figure*}[h!]
    \centering
    \includegraphics[trim = 0 0.2cm 0 2cm , clip=, width=0.45\columnwidth]{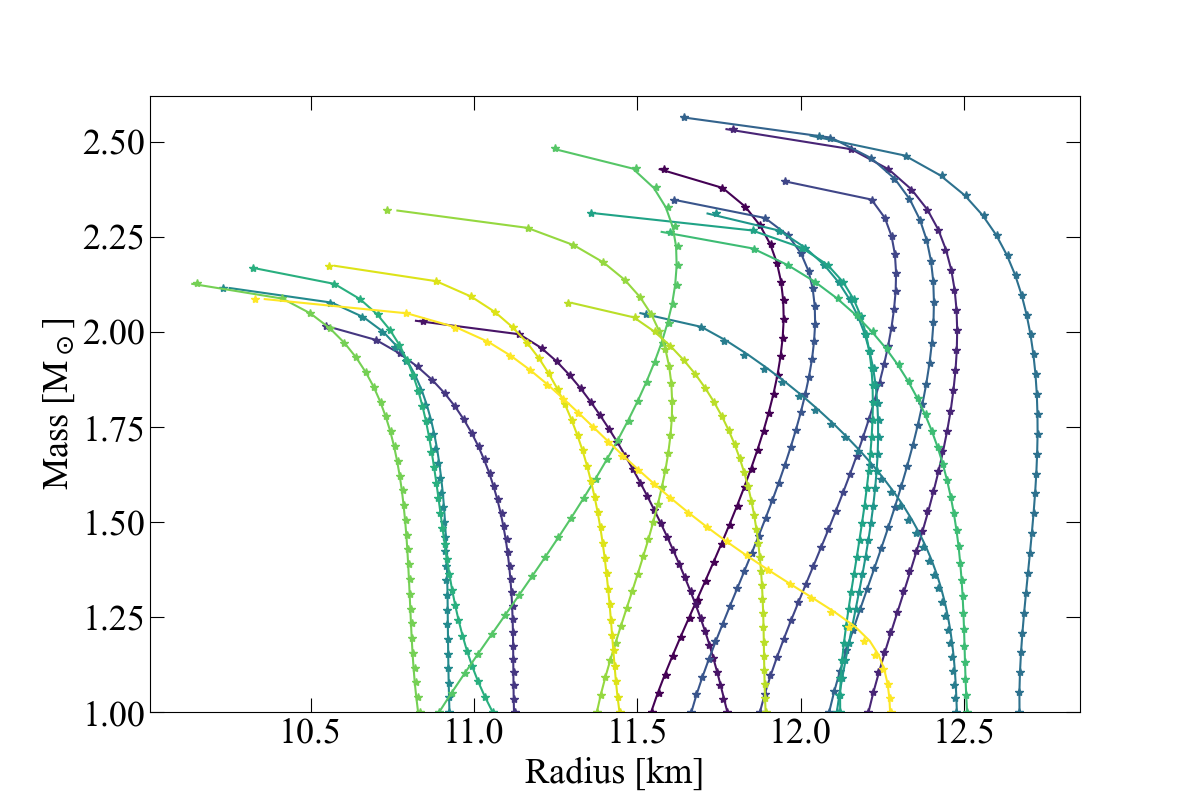}
    \includegraphics[trim = 0 0.2cm 0 2cm , clip=, width=0.45\columnwidth]{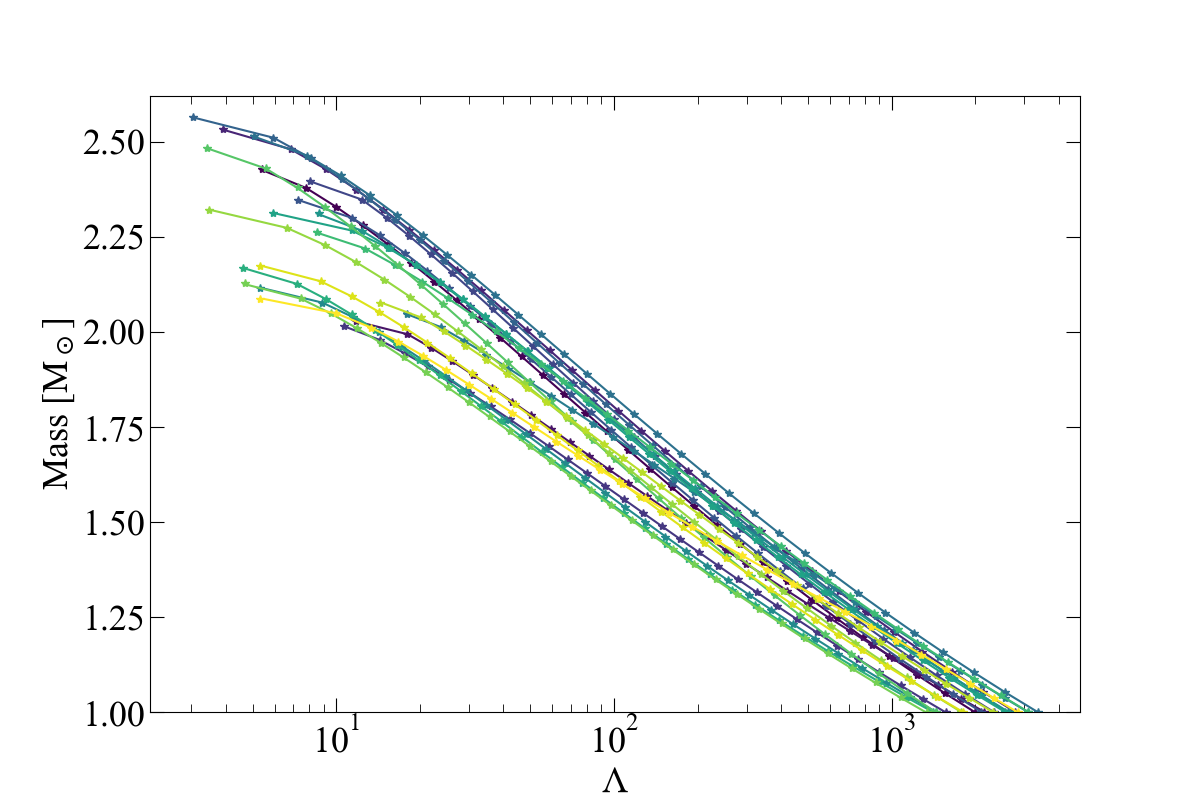}
    
    \caption{Examples of solutions to the TOV equations using the high-fidelity solver (lines) and the emulator predictions (points).
    Left: Mass--radius relations. Right: Mass--tidal-deformability curves.
    }
    \label{fig:radius_examples}
\end{figure*}

In Fig.~\ref{fig:radius_examples}, we show a set of example mass--radius and mass--tidal-deformability curves from the validation set. 
We find excellent agreement between the original high-fidelity EOS and the emulator output.

\section{LEC inference at every stage of the analysis}

\begin{figure*}[h!]
    \centering
    \includegraphics[height=0.28\columnwidth]{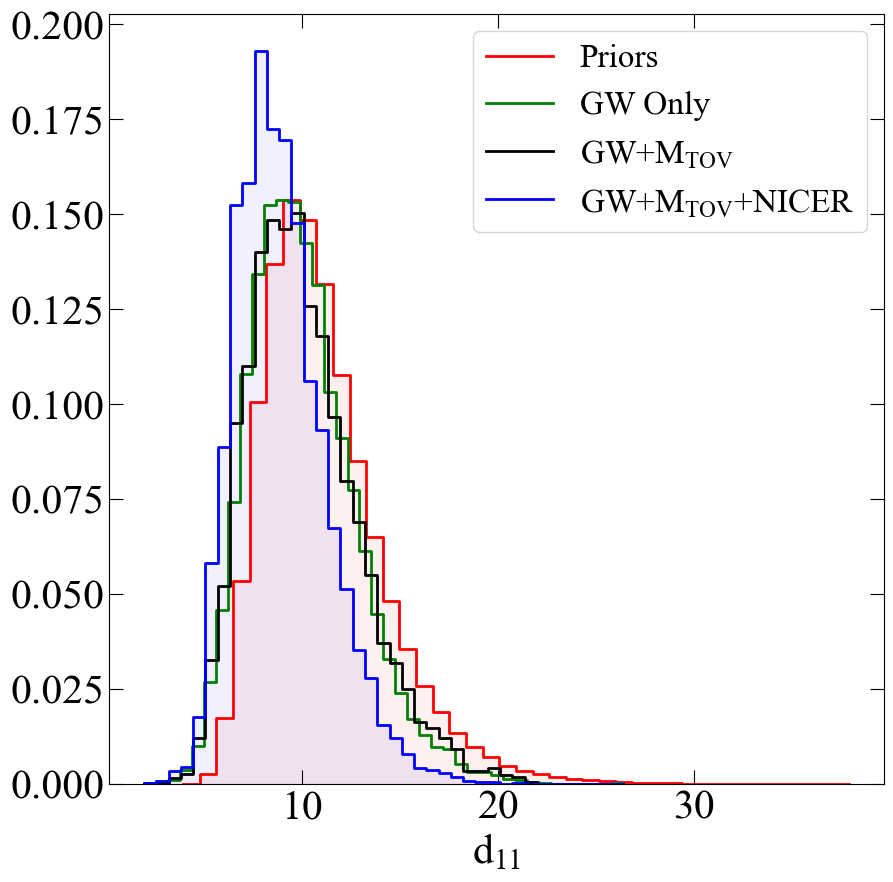}
    \includegraphics[height=0.28\columnwidth]{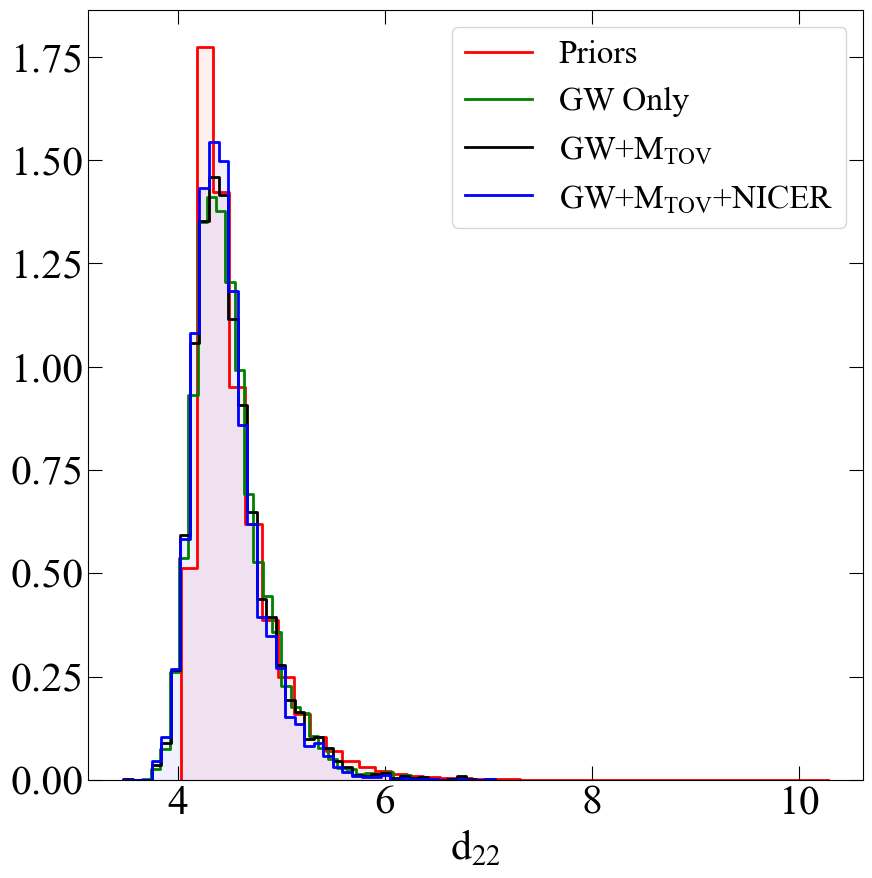}
    \includegraphics[height=0.28\columnwidth]{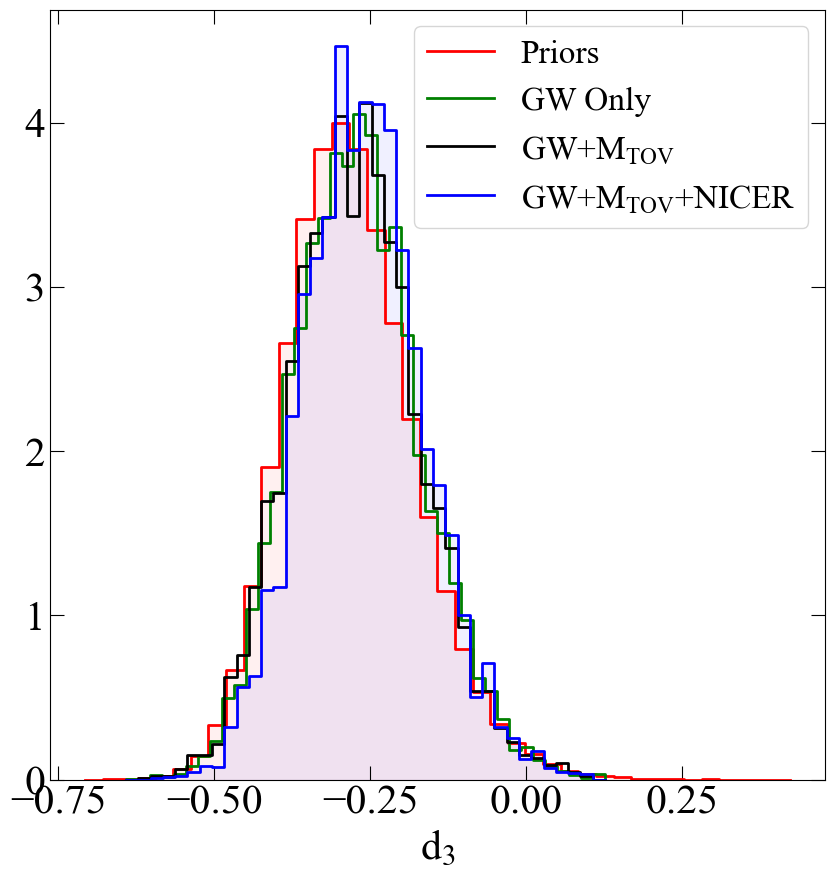}
    \includegraphics[height=0.28\columnwidth]{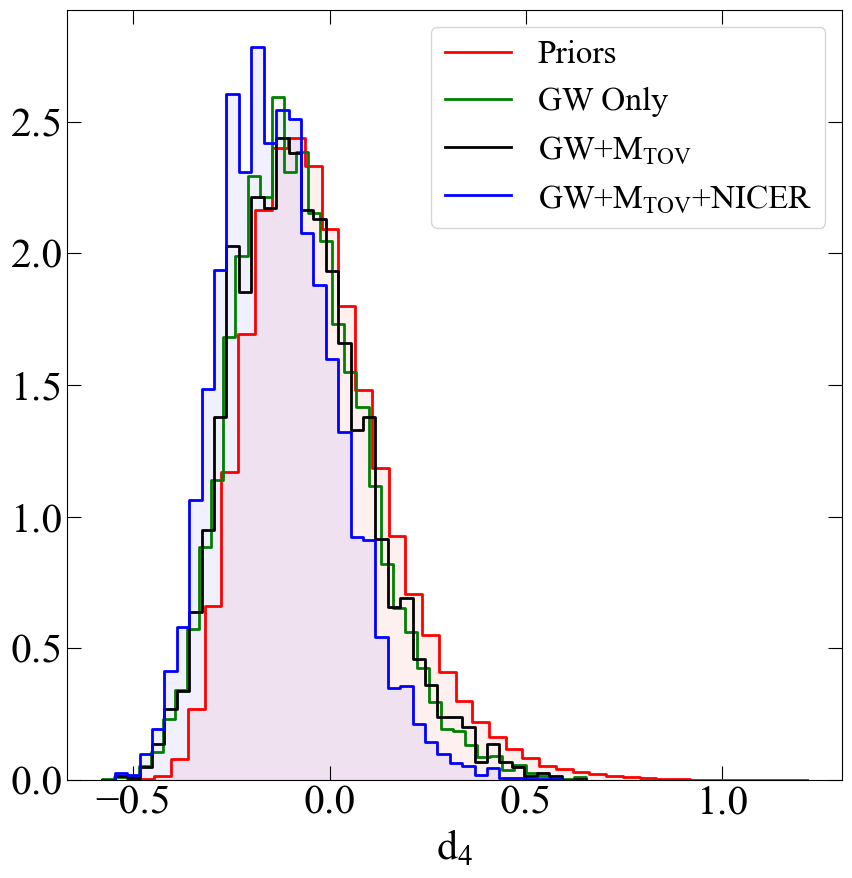}
    \includegraphics[height=0.28\columnwidth]{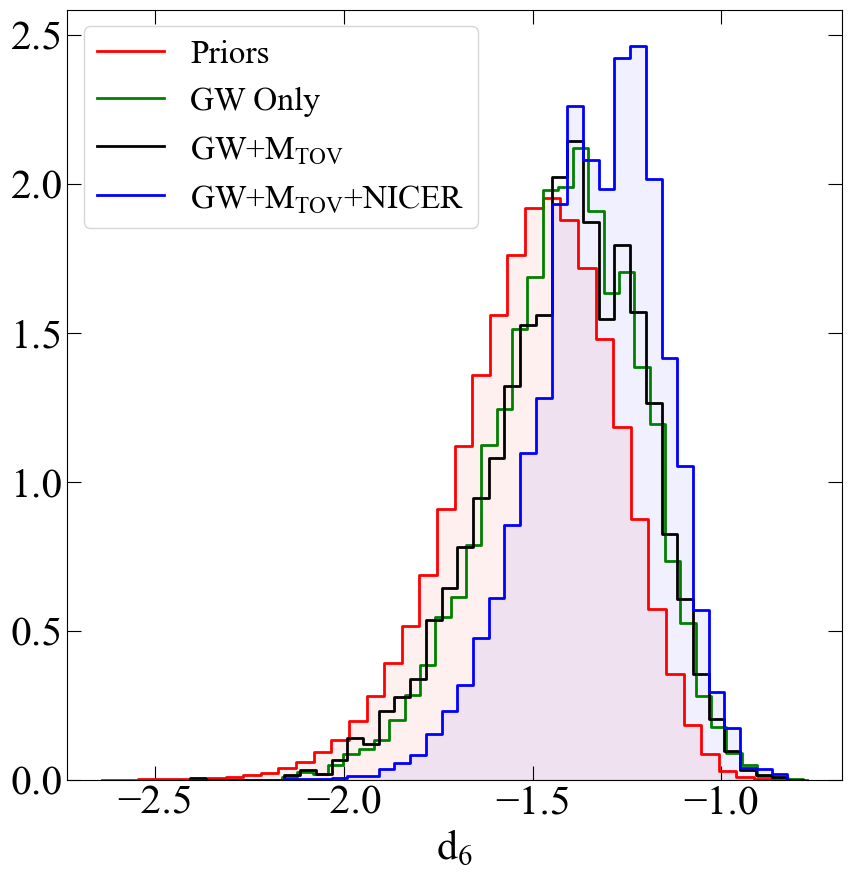}
    \includegraphics[height=0.28\columnwidth]{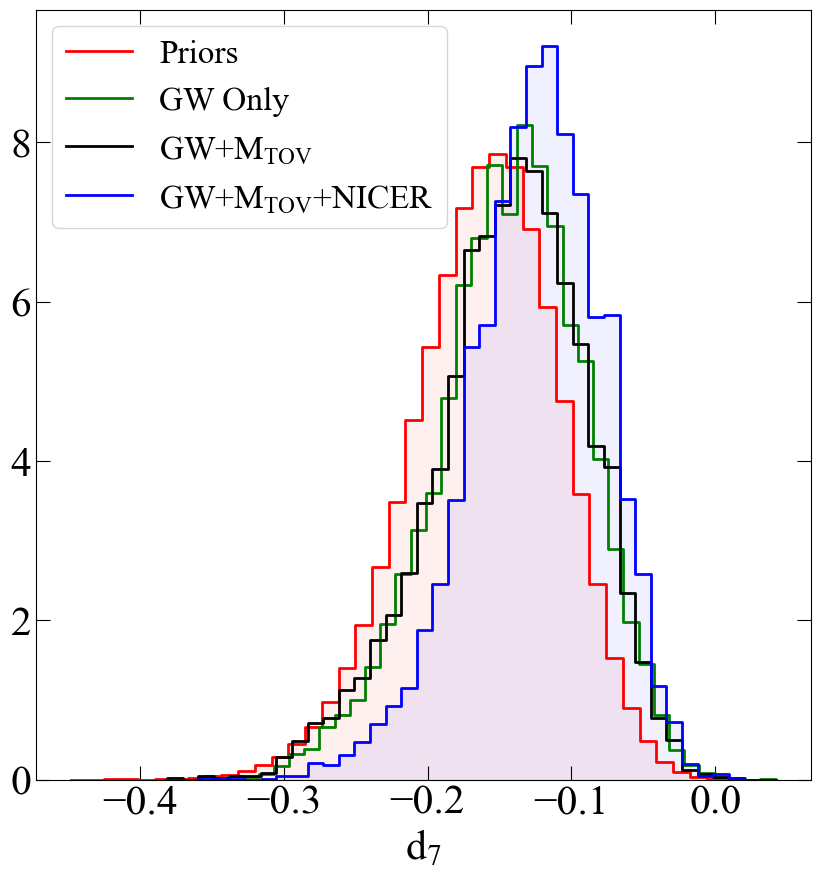}
    \caption{Priors (red) and posteriors at different stages in the analysis of astrophysical data for all six spectral LECs in neutron matter. 
    The priors are results of fits to NN scattering data. The posteriors include only GW data (green), GW and maximum-mass data (blue), and GW, maximum-mass, and NICER data (blue).
    }
    \label{fig:lecs_steps}
\end{figure*}

In Fig.~\ref{fig:lecs_steps}, we show the results of the LEC inference using current astrophysical data at each step in the analysis, so GW data only, GW and maximum mass data, and finally the full result shown in Fig.~\ref{fig:lecs} of the main manuscript.

\section{LEC inference using the Maryland NICER analysis}

\begin{figure*}[h!]
    \centering
    \includegraphics[height=0.28\columnwidth]{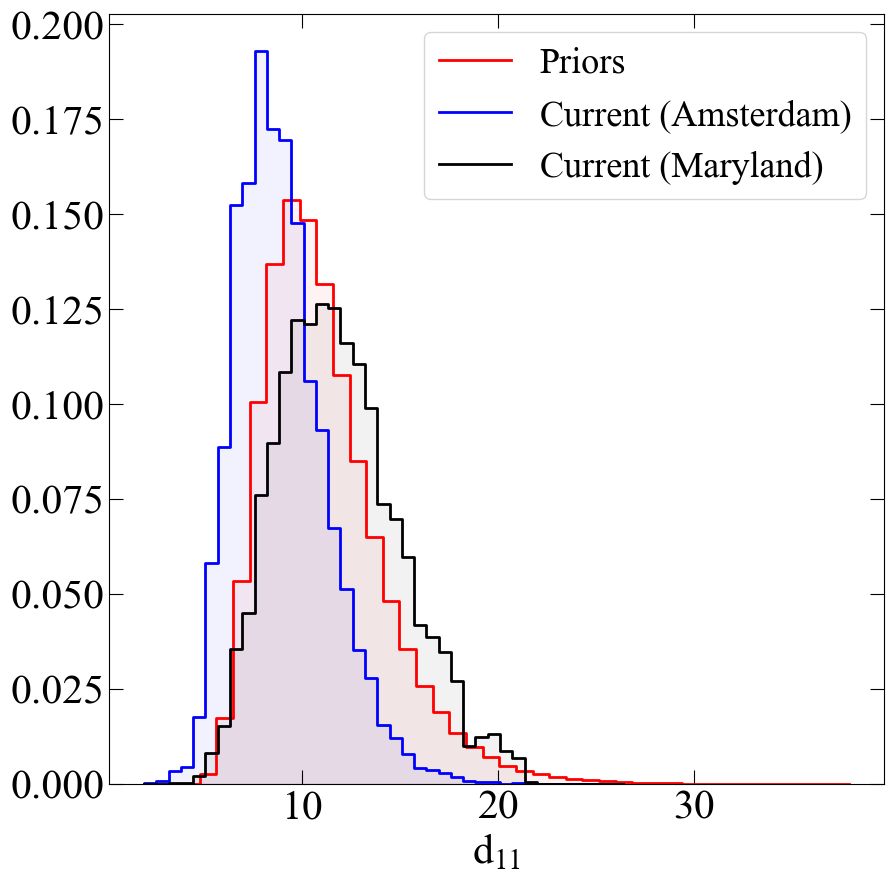}
    \includegraphics[height=0.28\columnwidth]{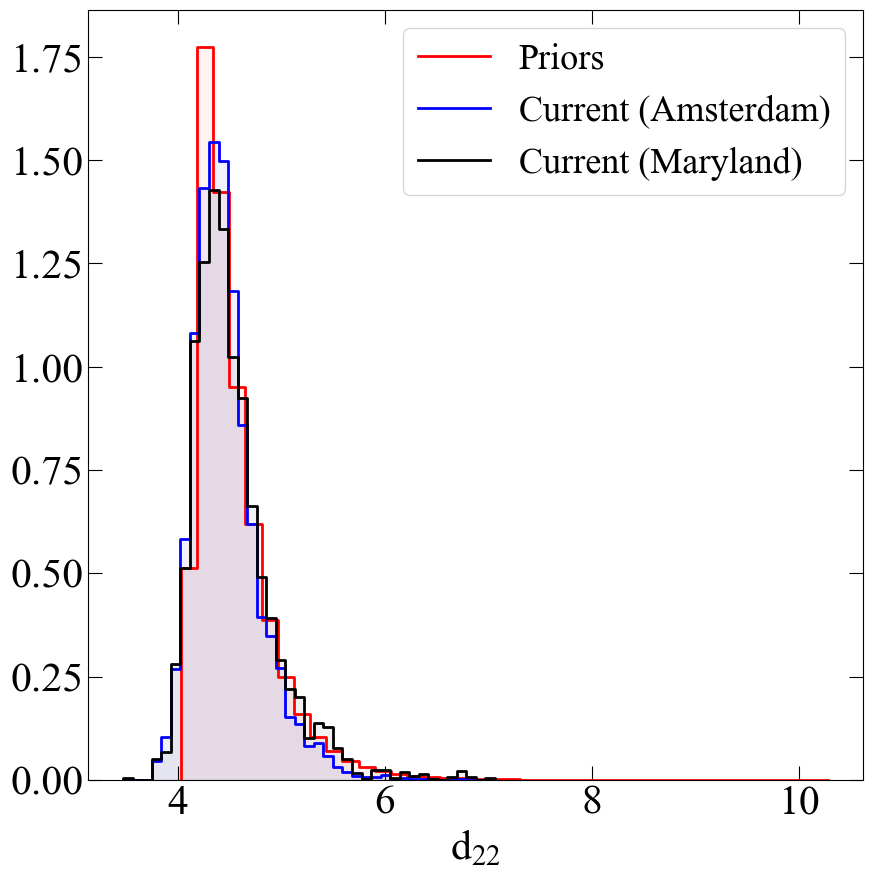}
    \includegraphics[height=0.28\columnwidth]{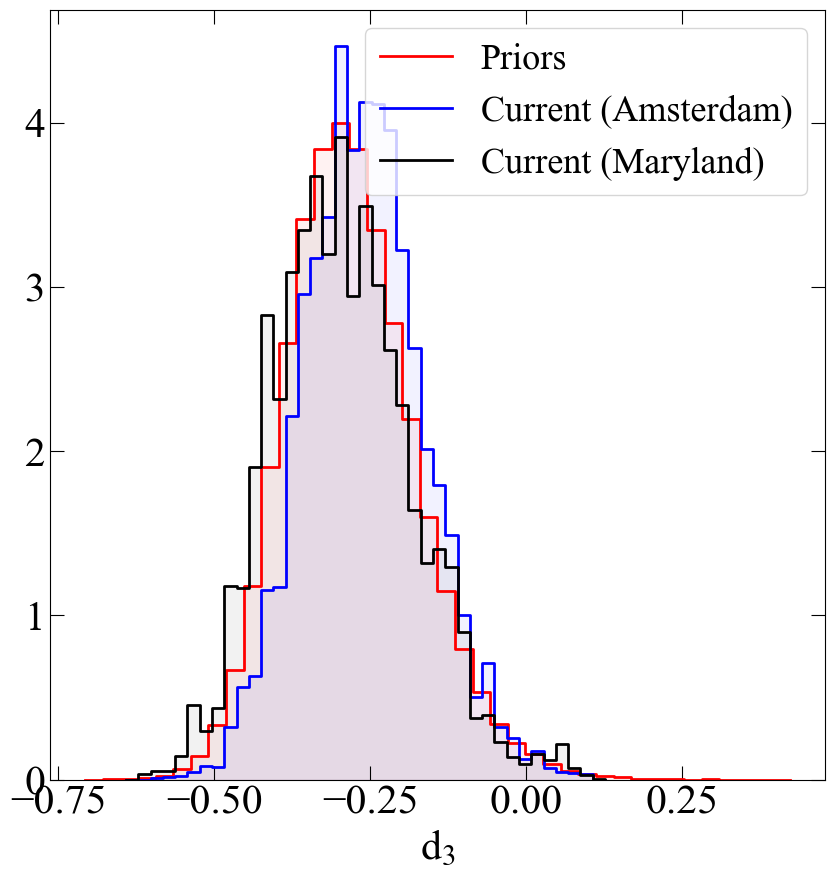}
    \includegraphics[height=0.28\columnwidth]{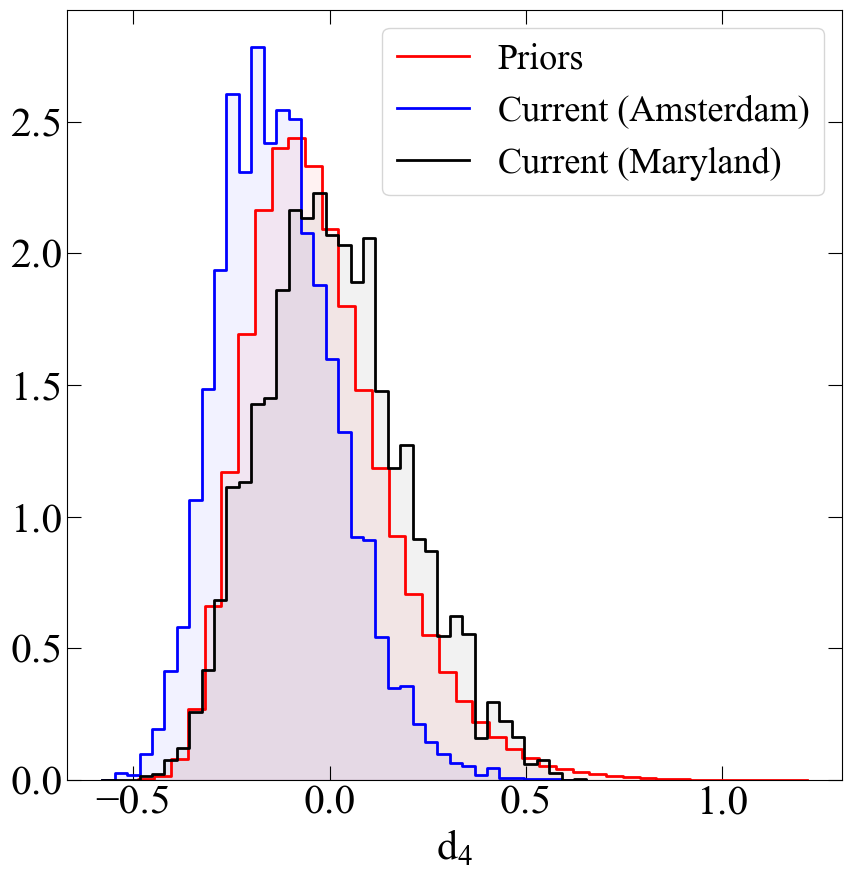}
    \includegraphics[height=0.28\columnwidth]{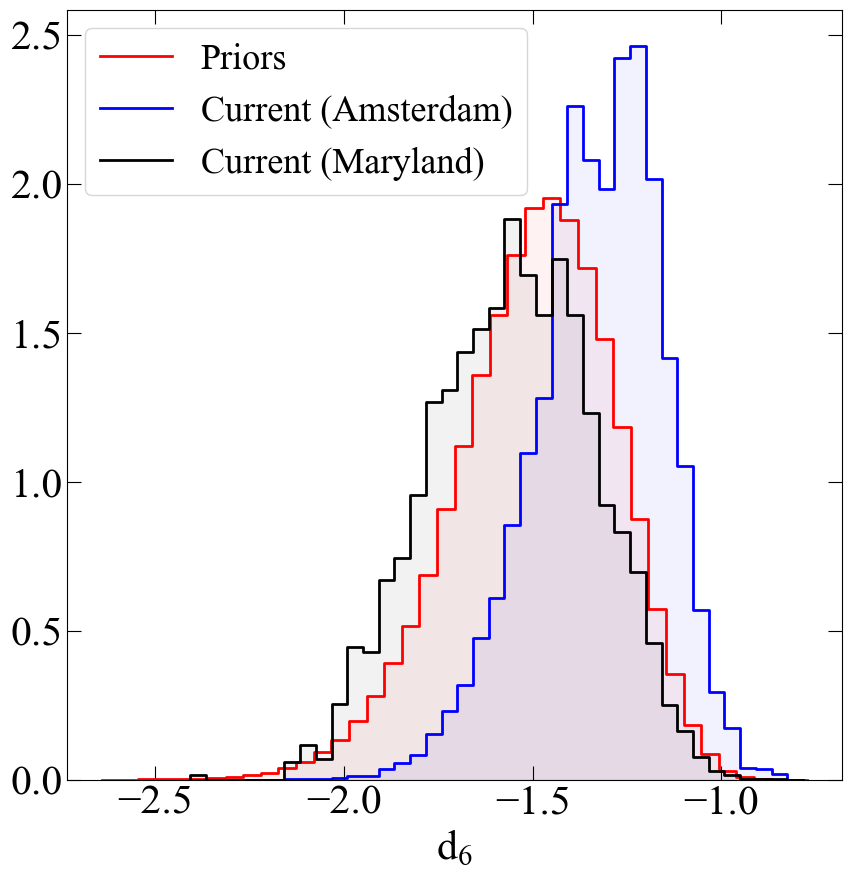}
    \includegraphics[height=0.28\columnwidth]{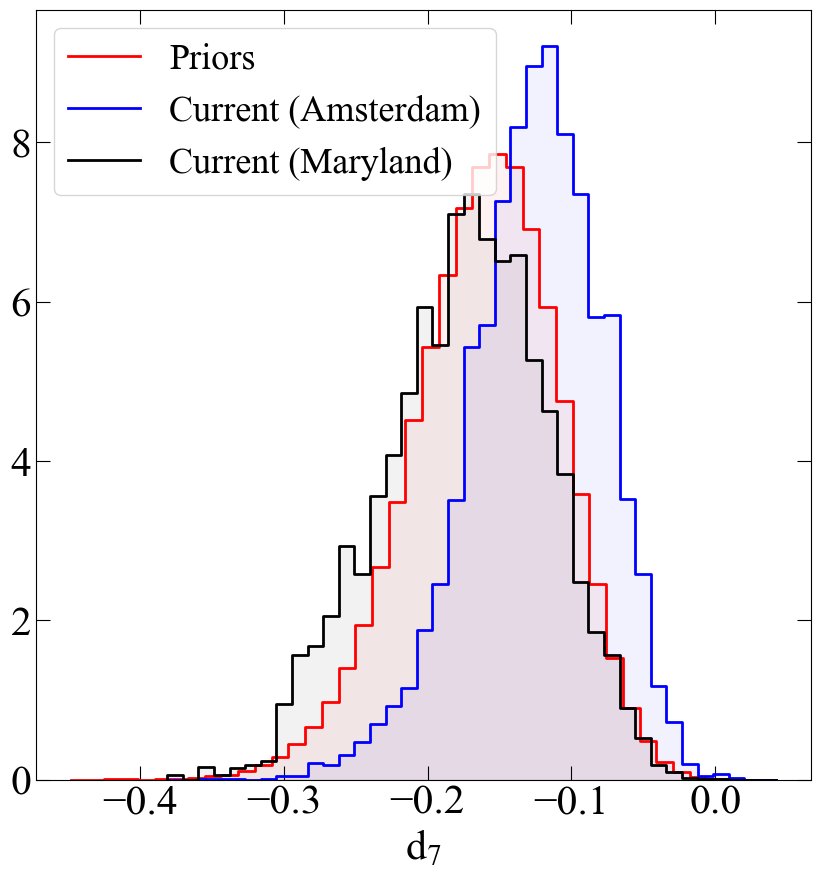}
    \caption{Priors (red) and posteriors for the analysis of current astrophysical data for all six spectral LECs in neutron matter when using GW, maximum-mass, and NICER data from the Amsterdam group (blue) and Maryland group (black).
    }
    \label{fig:lecs_Maryland}
\end{figure*}

In Fig.~\ref{fig:lecs_Maryland}, we show the posteriors for the spectral LECs if we replace the NICER results of the Amsterdam group with the three  published results of the Maryland group~\cite{Miller:2019cac,Miller:2021qha,Miller:2025qfq}.
We find moderate changes to the inferred posterior, but both posteriors are consistent within their uncertainties.

\newpage
\section{Correlation of LECs and NS radius}

\begin{figure*}[h!]
    \centering  \includegraphics[width=1.0\columnwidth]{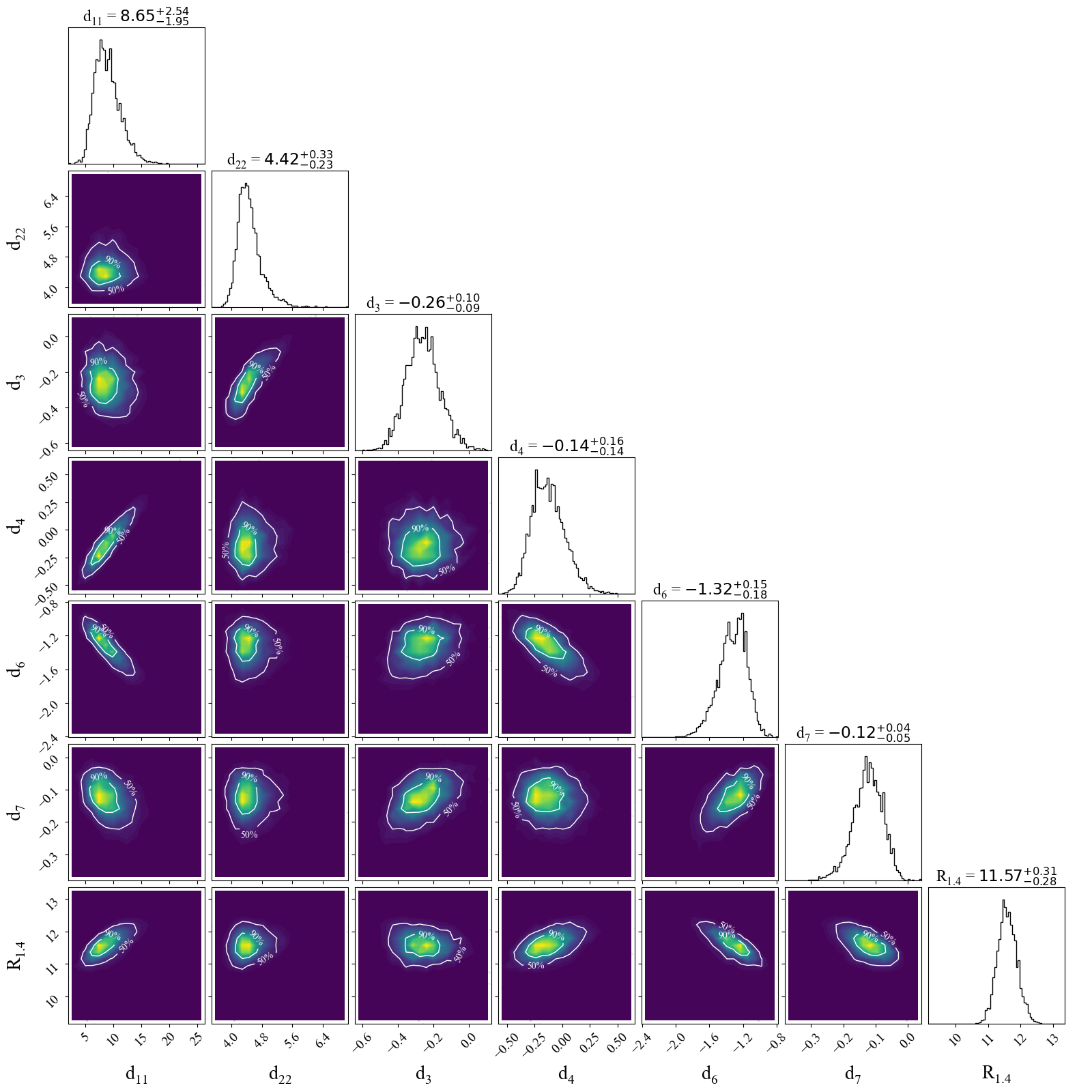}
    \caption{Corner plot for the posteriors of the 6 spectral LECs and the radius of a typical neutron star, R$_{1.4}$, using current astrophysical data.
    }
    \label{fig:corner_plot}
\end{figure*}

In Fig.~\ref{fig:corner_plot}, we present a corner plot relating the six spectral LECs to the radius of a typical $1.4$M$_{\rm sol}$ neutron star for our main result employing current astrophysical data.

\end{document}